\documentstyle[preprint,eqsecnum,aps,epsbox]{revtex}

\begin{document}
\draft
\preprint{KYUSHU-HET-29}
\title{A perturbative renormalization group approach \\
to light-front Hamiltonian
}
\author{Takanori Sugihara
\footnote{e-mail : sugi1scp@mbox.nc.kyushu-u.ac.jp}
}
\address{Department of Physics, Kyushu University,
Fukuoka 812-81, Japan}
\author{Masanobu Yahiro
\footnote{e-mail : yahiro@fish-u.ac.jp}
}
\address{National University of Fisheries, Shimonoseki 759-65, Japan}
\date{\today}
\maketitle
\vspace{-1.0cm}
\begin{abstract}
\hspace{1.0cm} A perturbative renormalization group (RG) scheme for
light-front Hamiltonian is formulated
on the basis of the Bloch-Horowitz effective Hamiltonian,
and applied to the simplest $\phi^4$ model
with spontaneous breaking of the $Z_2$ symmetry. RG equations
are derived at one-loop order for both symmetric and broken phases.
The equations are consistent with those calculated in
the covariant perturbation theory.
For the symmetric phase,
an initial cutoff Hamiltonian in the RG procedure
is made by excluding the zero mode
from the canonical Hamiltonian with an appropriate
regularization.
An initial cutoff Hamiltonian for
the broken phase is
constructed by shifting $\phi$ as $\phi \rightarrow
\phi-v$ in the initial Hamiltonian for the symmetric phase.
The shifted value $v$ is determined on a renormalization trajectory.
The minimum of the effective potential occurs on the trajectory.
\end{abstract}
\vfill\eject
\section{INTRODUCTION AND SUMMARY}
\label{sec:int}
Relativistic bound and scattering states of strongly interacting
particles are not understood well.
Quantum chromodynamics (QCD) is the typical case.
Such a problem is solved by constructing
a nonperturbative approach to relativistic quantum field theory.
Light-Front Tamm-Dancoff theory (LFTD) \cite{phw} is a hopeful
candidate for the approach. In LFTD, invariant masses of bound states
are obtained by diagonalizing a
light-front Hamiltonian after truncating the light-front Fock space.
The truncation is what is called the Tamm-Dancoff
approximation \cite{tam}. LFTD is precisely the Tamm-Dancoff
approximation applied to light-front field theory.
The approximation is believed to be reliable
at least for low-energy states.
This is really true
in two-dimensional gauge theories \cite{hara,sugi},
since the physical vacuum is trivial in the light-front field theory
\cite{my,ps}
and as the natural result the states may have simple structures.

The light-front field theory
raises complicated renormalization issues, when it is
applied to realistic four-dimensional field theories like QCD.
The light-front counterterms have complex structure and even
nonlocal in longitudinal direction, since infrared divergences
in longitudinal momentum and ultraviolet divergences
in transverse momentum arise simultaneously.
The structure is investigated for QED with perturbation
\cite{mpsw,atw}
and for the Yukawa model \cite{ghpsw} with an approximate
but nonperturbative manner.

As a powerful method for solving such complicated
renormalization issues in light-front
field theory, we consider Wilson's renormalization group (RG)
\cite{wk},
in which renormalization is achieved automatically
by finding a fixed point and a renormalization trajectory
( a flow running out of the point ).
Perry \cite{per1} applies
the Minkowski-space version \cite {wil1} of Wilson's RG
for light-front Hamiltonian. G{\l}azek and Wilson \cite{gw}
formulate a new perturbative RG scheme for Hamiltonian
by using a specially
designed similarity transformation. The two RG schemes
are based on perturbation and reliable for analyzes of RG flows
near a Gaussian fixed point.

The basic RG procedure for Hamiltonian is the following:
\begin{enumerate}
\item[(1)]
A bare Hamiltonian is regularized by truncating the Fock space
at a large cutoff energy $\Lambda_{0}$. The regularized Hamiltonian
$H_{\Lambda_{0}}$ is regarded as the initial Hamiltonian in the RG
procedure.
\item[(2)]
The truncated space is separated
into the lower- and higher-energy sectors.
An effective Hamiltonian $H_{\Lambda}$ for the lower-energy sector
is constructed
in a manner that it preserves physics of the lower-energy sector,
while the higher-energy sector is eliminated.
The cutoff is thus lowered to $\Lambda$.
In the actual derivation of
$H_{\Lambda}$, the finite transformation
($\Lambda_0 \rightarrow \Lambda$)
is expressed with successive small transformations
($\Lambda_0 \rightarrow \Lambda_1 \rightarrow \cdots \rightarrow
\Lambda$), and the $n$-th effective Hamiltonian
$H_{\Lambda_{n}}$ is derived with perturbation
from the $(n-1)$-th one $H_{\Lambda_{n-1}}$.

\item[(3)]
The cutoff $\Lambda$ is rescaled to $\Lambda_0$ by changing the energy
scale, and field variables are also rescaled so that
a fixed point may exit. In consequence,
$H_{\Lambda}$ is transformed into a new Hamiltonian $H_{\Lambda_{0}}'$
with the initial cutoff $\Lambda_{0}$.
The second and third steps progress from ultraviolet to infrared.
\end{enumerate}

The main purpose of this paper is to propose
a perturbative RG scheme which is more practical than those of Perry
and of G{\l}azek and Wilson. Our RG scheme differs from the two
schemes in the second step.
Perry uses the Bloch effective Hamiltonian \cite{b} which
contains an operator $R$ obeying a nonlinear equation
$
       R {\cal P}H_{\Lambda_{0}}{\cal P}
       - {\cal Q}H_{\Lambda_{0}}{\cal Q} R
       + R {\cal P}H_{\Lambda_{0}}{\cal Q} R
       - {\cal Q}H_{\Lambda_{0}}{\cal P} =0,
$
where ${\cal P}$ (${\cal Q}$) is a projector onto
the lower- (higher-) energy sector.
It is very hard to solve such a nonlinear equation without
perturbation.
Even if the equation is solved perturbatively by assuming
that all matrix elements of $R$ are small,
the solution has infinitely large matrix elements, since the matrix
elements contain vanishing energy differences in denominators.
Thus, perturbation does not work for the Bloch effective Hamiltonian.
The RG scheme of G{\l}azek and
Wilson also contains a nonlinear equation,
but the ``vanishing energy denominator" problem does not appear,
since it is designed so that
energy differences in denominators can be replaced by energy sums.
As it has not been applied to realistic field theories so far,
its application is highly expected. We then use
the Bloch-Horowitz \cite {bh} effective Hamiltonian
instead of the Bloch one.
This facilitates a perturbative RG procedure.
This RG scheme is applied to the simplest $\phi^4$ model
with spontaneous breaking of the discrete symmetry
$\phi \rightarrow -\phi$. Our results are summarized as follows.
\begin{enumerate}
\item[(i)]
Our RG scheme based on the Bloch-Horowitz effective Hamiltonian
is quite practical. The scheme is free from the ``vanishing
energy denominator" problem, so that RG equations are easily derived
with perturbation. The resultant RG flows depend on
eigenenergies $E_i$ of
$H_{\Lambda_{0}}$, since so does the effective Hamiltonian.
The dependence is negligible for $\Lambda$ much
larger than $E_i$. The state dependence becomes significant
as $\Lambda$ decreases,
but it does not make any trouble.
Renormalization is achieved by finding a renormalization trajectory
for the lowest state with $E_1$. Renormalization trajectories for
higher energy states are obtainable with LFTD
from that for the lowest state.
The renormalization trajectories, each with different
$E_i$, converge on a fixed point as $\Lambda$ goes to infinity.
In principle, this RG scheme is applicable
not only for light-front Hamiltonians but also for equal-time
Hamiltonians.
\item[(ii)]
Renormalization group (RG) equations for mass $\mu$ and
coupling $\lambda$ are derived at one-loop order, where
all irrelevant operators generated by RG transformation are
removed as a reasonable approximation.
The invariant mass regularization \cite{per1}
is adopted in this paper,
but it breaks covariance and cluster property, so
the running mass and coupling constant depend
on momenta of spectators if they exist.  The dependence
is, however, very weak, as long as
$\Lambda $ is at least several times
larger than $\mu$ and $M_i$ which are assumed to be of order
the physical mass scale $\Lambda_{phys}$.
So it is neglected as a reasonable approximation in this paper.
The regularization also excludes the zero mode
(a mode with zero longitudinal momentum) from
the canonical Hamiltonian.
The zero mode is responsible for spontaneous symmetry breaking,
that is, the order parameter $\langle 0| \phi |0 \rangle$
for the $Z_2$ symmetry never becomes nonzero without the mode.
This means that the RG equations calculated with the initial cutoff
Hamiltonian are
correct only for the symmetric phase. In fact, a flow diagram drawn
with the RG equations shows not only that two phases exist,
but also that
tachyons come out in the broken phase.
\item[(iii)]
In light-front field theory, Hamiltonians are different between the
two phases, while their vacua are always trivial.
Result (ii) indicates
that for the symmetric phase the initial Hamiltonian
 $H_{\Lambda_{0}}$ is obtained just by
removing the zero mode from the canonical Hamiltonian with an
appropriate regularization. A problem is how to construct another
$H_{\Lambda_{0}}$ valid for the broken phase. Once the zero mode is
switched off, the system has to sit in the bottom of the effective
potential. We then have to shift
$\phi$ as $\phi \rightarrow \phi-v$ in the initial Hamiltonian
for the symmetric phase.
Once a renormalization trajectory is found, $v$
is determined as a function of $\Lambda$ on the trajectory.
\item[(iv)]
The RG equations for
$\mu$, $\lambda$ and $v$ calculated in the present framework
are compared with those in
the covariant perturbation theory. For $\Lambda \gg \Lambda_{phys}$,
both agree with each other, except for the following point.
In the covariant theory
a contribution of the tadpole (Fig. 1) is explicitly calculated
and then present in the RG equations. On the other hand,
the contribution is removed in our normal ordered
$H_{\Lambda_{0}}$,
so it does not appear in our RG equations explicitly.
\item[(v)]
At the one-loop level,
we find by using the present RG equations that
$\mu(\Lambda)^2 + \lambda(\Lambda)v(\Lambda)^2/6$ is invariant for any
$\Lambda$ and by calculating the effective
potential that the RG invariant quantity is zero
when the system sits in
the bottom of the potential. The resultant relation
$\mu(\Lambda)^2 + \lambda(\Lambda)v(\Lambda)^2/6 =0$
are thus a condition for the
system to be in the bottom, and the renormalization trajectory
should satisfy the relation. This is confirmed. A way of
finding the relation without calculating the effective potential is
furthermore presented.
\end{enumerate}
Throughout all the results, we conclude that the present perturbative
RG scheme is quite practical and valid at least near a Gaussian fixed
point.
This method is applicable for both the symmetric and broken phases.
Application of the present method to QCD may not be straightforward,
since it is known in the equal-time field theory that
the QCD vacuum is much more complicated than that of the present
model. Further study on how to construct $H_{\Lambda_0}$ in the QCD
case is highly expected.

Section \ref{sec:rgs} presents our RG scheme and shows result (i).
In section \ref{sec:phi}, the scheme is applied to $\phi^4$ model
in 3+1 dimensions with spontaneous breaking of the $Z_2$ symmetry.
In subsections \ref{rg1} and \ref{rg23}, we
derive RG equations for the symmetric phase to show result (ii).
Subsection \ref{rga} considers how to derive RG equations for the
broken phase and shows result (iii) and a part of result (v).
In subsection \ref{effpot},
we compare the RG equations with those calculated
in the covariant perturbation theory and show results (iv) and (v).
In Appendix A we evaluate loop integrals present in RG equations
and show a part of result (ii).

\section{RENORMALIZATION GROUP SCHEME}
\label{sec:rgs}

A RG scheme is proposed along three steps (1)-(3) mentioned
in section \ref{sec:int}. The scalar field theory is considered
as an example.

As step (1), the light-front bare Hamiltonian is regularized with a
boost-invariant regularization. As a feature, the bare Hamiltonian
has no coupling between center-of-mass and intrinsic motions,
indicating that the two types of motions are
independent of each other.
The regularization keeps the property.
Among some possible boost-invariant
regularizations, we take the invariant mass
regularization \cite{per1}, since some loop integrals
appearing in the RG equations are analytically calculable.

In light-front kinematics
a free particle with longitudinal and transverse
momenta, ${\bf k} \equiv (k^+, {\bf k}_\perp)$ , has an energy
$\epsilon_{\bf k} \equiv ({{\bf k}_\perp}^2+\mu^2)/(2k^+)$.
An invariant mass $M$ of an $n$-body Fock state,
each particle with a momentum ${\bf k}_i$, is then defined as
$
               E = ({{\bf P}_\perp}^2+M^2)/(2P^+)
$
for the total energy
$
E = \sum_{i=1}^{n} \epsilon_{{\bf k}_i}
$
and the total momentum
$
{\bf P} \equiv (P^+, {\bf P}_\perp)
=\sum_{i=1}^{n} {\bf k}_i$.
The invariant mass regularization excludes
all Fock states with $M$ larger than an initial  cutoff
$\Lambda_{0}$. Since $M$ diverges when $k^+=0$,
the mode with $k^+=0$ is removed here.
The initial cutoff Hamiltonian is denoted by $H_{\Lambda_{0}}$.

As step (2),
the truncated Fock space is cut at $M=\Lambda$ smaller than
$\Lambda_{0}$ by a finite amount, and separated
into the lower and higher $M$ sectors, i.e.
the ${\cal P}$ and ${\cal Q}$ sectors.
We use the Bloch-Horowitz effective Hamiltonian
\begin{equation}
       H_{\Lambda}(M_i)
       = {\cal P}H_{\Lambda_{0}}{\cal P}
        +{\cal P}V{\cal Q}
          \frac{1}{E_i-{\cal Q}H_{\Lambda_{0}}{\cal Q}}
         {\cal Q}V{\cal P},
  \label{effh}
\end{equation}
where $H_{\Lambda_{0}}$ is composed of
the free and interaction parts, $H_0$ and $V$.
The effective Hamiltonian has been derived with the projectors
${\cal P}$ and ${\cal Q}$ from
the Schr\"odinger equation
$H_{\Lambda_{0}}|\Psi_i\rangle=E_i|\Psi_i\rangle$,
where the $i$-th eigenmass $M_i$ is determined from $E_i$ with the
dispersion relation
$
               E_i = ({{\bf P}_\perp}^2+M_i^2)/(2P^+)
$.
Eigenvalues and eigenstates, $E'_i$ and $|\Psi'_i\rangle$,
of $H_{\Lambda}$ satisfy
$E'_i=E_i$ and $|\Psi'_i\rangle={\cal P}| \Psi_i \rangle$.
The effective Hamiltonian thus preserves physics
of the ${\cal P}$ sector.

In principle, the finite transformed Hamiltonian $H_{\Lambda}$ is
derivable with (\ref{effh}) from $H_{\Lambda_0}$. In practice,
however,
the finite transformation ($\Lambda_0 \rightarrow \Lambda$) is
described with successive small transformations
($\Lambda_0 \rightarrow \Lambda_1 \rightarrow \ldots \rightarrow
\Lambda$), and the $n$-th Hamiltonian $H_{\Lambda_n}$ is derived from
the $(n-1)$-th one with the Bloch-Horowitz effective
Hamiltonian formalism. This prescription has a merit;
see subsection \ref {rg23}.
The $n$-th effective Hamiltonian $H_{\Lambda_n}$
includes the ${\cal Q}$-space Green function
$G \equiv 1/(E_i-{\cal Q}H_{\Lambda_{n-1}}{\cal Q})$,
where the ${\cal Q}$ space is $\Lambda_{n} < M \le \Lambda_{n-1}$ and
the ${\cal P}$ space is $ M \le \Lambda_{n}$.
The Green function is related to the free one
$G_{\rm 0} \equiv 1/(E_i-{\cal Q}H_{\rm 0}{\cal Q})$ as
\begin{equation}
      G
      =
      G_{\rm 0}+
      G_{\rm 0}{\cal Q} V {\cal Q}G,
\label{g}
\end{equation}
where use has been made of the identity
$A^{-1}-B^{-1}=B^{-1}(B-A)A^{-1}$. The full Green function $G$
is obtained by solving (\ref {g}) perturbatively. When $E_i$
is above the two-body threshold energy
$E_{thresh}= \epsilon_{{\bf k}_1}+\epsilon_{{\bf k}_2}$, we must
solve a scattering problem by replacing $E_i$ by $E_i+ i\delta $.

The RG procedure is completed by the scale transformation
of Step (3).
The effective Hamiltonian $H_{\Lambda}$ is transformed into
a new Hamiltonian $H_{\Lambda_{0}}'$ with a cutoff $\Lambda_{0}$
by scaling transverse momenta ${\bf k}_\perp$ as
\begin{eqnarray}
     {\bf k}_\perp&=&\frac{\Lambda}{\Lambda_{0}}{\bf k}_{\perp}',
\end{eqnarray}
since the ${\cal P}$ space ($M<\Lambda$) is expanded to
the ${\cal P}+{\cal Q}$ space ($M<\Lambda_{0}$)
by the transformation; precisely speaking,
a total energy of an $n$-body Fock state belonging to
the ${\cal P}$ space is varied in the region
\begin{equation}
     0 < \sum_{i=1}^{n}
         \frac{\mu^2+{\bf k}_{\perp i}^2}{2k^+_i}
       < \frac{\Lambda^2+{\bf P}_{\perp}^2}{2P^+} ,
\label{scale:p}
\end{equation}
and the scaling expands the region to
\begin{equation}
     0 < \sum_{i=1}^{n}
         \frac{(\Lambda_{0}/\Lambda)^2\mu^2+{\bf k}_{\perp i}'^2}
              {2k^+_i}
       < \frac{\Lambda_{0}^2+{\bf P}_{\perp}^{'2}}{2P^+}.
\end{equation}
In addition to the transverse momenta,
the field variables $\phi({\bf k})$ ( the creation and annihilation
operators, $a^\dagger({\bf k})$ and $a({\bf k})$ )
and the Hamiltonian $H_{\Lambda}$
are also scaled to \cite{per1}
\begin{eqnarray}
     \phi(k^+,{\bf k}_\perp)&=&\zeta \phi'(k^+,{\bf k}_{\perp}'),
\nonumber\\
     H_{\Lambda}(\phi(k^+,{\bf k}_\perp))
     &=&
     \eta^{-1} H_{\Lambda_{0}}'(\phi'(k^+,{\bf k}_{\perp}')).
\label{scale:h}
\end{eqnarray}
The constants $\zeta$ and $\eta$ are determined
so that fixed points can exist. There exists a
Gaussian fixed point in four dimensional $\phi^4$ model.
At the fixed point, $H_{\Lambda_{0}}$
is reduced to the free part $H_0(\Lambda_{0};\phi(k^+,{\bf k}_\perp))$,
so the Bloch-Horowitz effective Hamiltonian $H_{\Lambda}$
and the rescaled one $H_{\Lambda_{0}}'$ are easily
derived as
$H_{\Lambda}={\cal P}H_0(\Lambda_{0};\phi(k^+,{\bf k}_\perp)){\cal P}$
and
$H_{\Lambda_{0}}'
=\eta H_0(\Lambda_{0};\zeta\phi'(k^+,{\bf k}_\perp'))$.
The constants $\zeta$ and $\eta$ are determined from the condition that
$H_{\Lambda_{0}}'=H_{\Lambda_{0}}$. The constants thus obtained are
\cite{per1}
\begin{equation}
     \zeta=(\Lambda/\Lambda_{0})^{-1}, \hskip 10pt
     \eta =(\Lambda/\Lambda_{0})^{-2} .
\end{equation}

Adopting the Bloch-Horowitz effective Hamiltonian in step (2)
makes the present RG procedure practical, because we can easily
solve Eq.(\ref{g}) with perturbation. For $E_i < E_{thresh}$,
the solution contain energy differences $E_i-E$ in denominators, where
$
               E = ({{\bf P}_\perp}^2+M^2)/(2P^+)
$
and $M$ belongs to the ${\cal Q}$ space. The denominators
never vanish, because $E \ge E_{thresh}$ for any $\Lambda$ and
there exists an energy gap between $E_i$ and $E_{thresh}$.
The present RG scheme is thus free from
the ``vanishing energy denominator" problem.

The effective Hamiltonian is not Hermitian, since
it depends on $M_i$.
However, it does not make any trouble. An only
difference from the ordinary RG
is that the running mass $\mu$ and the running coupling $\lambda$
depend on not only $\Lambda$ but also $M_i$.
The $M_i$ dependence or the state dependence becomes negligible
for $ \Lambda \gg M_i$,
since on the right hand side of (\ref{g})
      $E_i-E=(M_i-M)/(2P^+) \approx -M/(2P^+)$ as a result of
$M_i \ll \Lambda < M < \Lambda_{0}$.
We are interested in low-energy physics, so $M_i$ is considered to be
of order $\Lambda_{phys}$.
The parameters $\mu$ and $\lambda$ thus has no $M_i$ dependence
for $\Lambda \gg \Lambda_{phys}$.

When $\Lambda$ is of order $\Lambda_{phys}$, the $M_i$ dependences of
$\mu$ and $\lambda$ become significant.
For any scattering state ( with $E_i > E_{thresh}$ ),
$M_i$ is just an input parameter determined
from the initial condition of scattering,
so one can draw a RG flow. For bound states
( with $E_i < E_{thresh}$ ),
on the contrary, $M_i$'s are unknown. Only an exception
is the lowest mass $M_1$, since it is set to
the physical mass $M_{phys}$ as a renormalization condition.
Other $M_i$'s each are determined in a prescription mentioned below.

Our renormalization procedure starts with drawing a flow diagram
for the lowest state by solving RG equations under the condition
$M_1=M_{phys}$. It essentially ends up
with finding out a renormalization trajectory in the flow diagram,
since
renormalization trajectories for excited states
are obtainable from that
for the lowest state just by replacing $M_1$ by $M_i$,
if $M_i$ is given; this is obvious, if one knows
analytic solution to RG equations.
All the trajectories converge at a fixed point
as $\Lambda$ goes to infinity.
The $M_i$'s are given for scattering states, but not
for bound states. The invariant masses of bound states are obtained
as follows. Suppose that the trajectory for the lowest state is found.
We choice a point A on the trajectory
at which $\Lambda$ is much larger than
$\Lambda_{phys}$; the cutoff is denoted by $\Lambda_A$.
If one can solve the Schr\"odinger equation
$
\{ H_{\Lambda_A}(M_1)-({{\bf P}_\perp}^2+M_i^{'2})/(2P^+) \}
|\Psi'_i\rangle =0
$
with LFTD, the approximate
eigenvalues $M_i^{'2}$ each contain errors of
order $(M_i^2 - M_1^2)/\Lambda_{A}^2$.  The errors are thus
negligible for low-lying states. For such large Fock space,
LFTD may not be so practical,
since it demands a large number of basis functions to
diagonalize the Hamiltonian accurately.
If one comes across this problem
in practical calculations, take
the following self-consistent way as a second choice.
First draws a RG flow starting from point A
with an initial estimate $M_i^{(1)}$ of $M'_i$.
Next choose a point B on the flow at which $\Lambda$ is of order
$\Lambda_{phys}$.
For such small $\Lambda$ LFTD is a powerful tool, unless $\lambda$
is very large. At point B one can then diagonalize
the effective Hamiltonian to get
the second estimate $M_i^{(2)}$. Again, draw a flow running out of
point A with $M_i^{(2)}$, and so on.
This procedure progresses until we get $M_i^{(n+1)}=M_i^{(n)}$.
The mass thus obtained is equal to the approximate mass $M'_i$.

\section{$\phi^4$ MODEL}
\label{sec:phi}
\subsection{The first step of RG procedure}
\label{rg1}

Our convention for light-front coordinates is
$x^{\pm}=(x_0\pm x_3)/\sqrt{2}$ and
$x^i_\perp \equiv x^i (i=1,2)$. The quantity
$x^+$ is chosen as the "time" direction
along which all the states are evolved, so $x^-$ and $x_{\perp}$
become the longitudinal and transverse directions.
The metric tensor is then
given by
$g^{+-}=g^{-+}=g_{+-}=g_{-+}=1, g^{ii}=g_{ii}=-1 (i=1,2)$
and zero for others.

The Lagrangian density of $\phi^4$ model in $4$ dimensions is
\begin{eqnarray}
    {\cal L} &=& \frac{1}{2}\partial_{\mu}\phi\partial^{\mu}\phi
                -\frac{\mu^2}{2}\phi^2 -\frac{\lambda}{4!}\phi^4
\end{eqnarray}
for a real scalar field $\phi$. The commutation relation for the field
derived with the Schwinger's action principle \cite{chang} is
\begin{equation}
       [\phi(x),\partial_{-}\phi(y)]_{x^+ = y^+}
       =\frac{i}{2} \delta^{3}({\bf x}-{\bf y}).
\label{comm}
\end{equation}
The field is expanded in terms of free waves at $x^+=0$,
\begin{equation}
       \phi({\bf x})
       =
       \int \frac{d^{3}{\bf k}}{\sqrt{(2\pi)^{3}2k^+}}
            [a          ({\bf k})e^{-ik^+x^-
                                    +i{\bf k}_\perp \cdot
                                      {\bf x}_\perp
                                   }
         +
             a^{\dagger}({\bf k})e^{ ik^+x^-
                                    -i{\bf k}_\perp \cdot
                                      {\bf x}_\perp
                                   }
            ],
\label{exp}
\end{equation}
where
\begin{equation}
       \int d^{3}{\bf k}
       \equiv
       \int_{      0}^{+\infty} dk^+
       \int_{-\infty}^{+\infty} d{\bf k}_{\perp}.
\end{equation}
Inserting this form into (\ref{comm}) yields a relation
\begin{equation}
       [a({\bf k}),a^\dagger({\bf k}')]
       =\delta^{3}({\bf k}-{\bf k}')
\end{equation}
between the coefficients of expansion for positive $k^+$ and $k'^+$.
Obviously, $a({\bf k})$ and $a^\dagger({\bf k})$  with
positive $k^{+}$ are
annihilation and creation operators for the Fock vacuum.
As an important feature of light-front field theory, furthermore,
they are annihilation and creation operators also for the true
vacuum $|0\rangle$, since it is proven that
$a({\bf k}) |0\rangle =0$ for positive $k^+$ \cite{my}.
This indicates that the true
vacuum is trivial in light-front field theory,
at least as far as the zero mode is neglected; it is still
true even after the zero mode is included explicitly with an
appropriate way \cite{ps}.
The Fock space is then constructed
by acting the creation operator on the true
vacuum: $ |{\bf k}_1,{\bf k}_2,\dots,{\bf k}_n \rangle
       \equiv
       \prod_{i=1}^{n}a^\dagger({\bf k}_i) |0\rangle .$

The Hamiltonian derived from the energy-momentum tensor \cite{chang}
is
\begin{equation}
    H = \int d^3 x
           \bigg \{
                 \frac{1}{2}\Big[
                 \sum_{i=1}^{2}(\partial_i \phi)^2+
                 \mu^2 \phi^2]+\frac{\lambda}{4!}:{\phi^4}: \bigg \},
\end{equation}
where we have removed the tadpole (Fig. 1) by normal ordering
the interaction term. The Hamiltonian is rewritten into
\begin{equation}
      H_{\Lambda_0}
      =
      H_{\rm 0} + V ,
\label{hlambda0}
\end{equation}
\begin{eqnarray}
      H_{\rm 0}
      &=&
      \sum_{n=1}\frac{1}{n!}
      \int\Big[\prod_{i=1}^{n}d{\bf k}_i\Big]
      \theta(\epsilon_{\Lambda_0}-\sum_{i=1}^{n}\epsilon_{{\bf k}_i})
      \bigg[ \sum_{i=1}^{n}\epsilon_{{\bf k}_i}
      \bigg]
\nonumber\\
      & &
      \hskip 50pt \times
      |{\bf k}_1,{\bf k}_2,\dots,{\bf k}_i\rangle
      \langle{\bf k}_1,{\bf k}_2,\dots,{\bf k}_i|,
\\
      V
      &=&
      \sum_{n=2}v_{n,n} +
      \sum_{n=1}(v_{n+2,n} + v_{n,n+2}) ,
\end{eqnarray}
where
\begin{eqnarray}
      v_{n,n}
      &\equiv&
      \frac{3\bar \lambda}{2(n-2)!}
      \int
           \Big[
           \frac{ \prod_{i=1}^{n} d{\bf k}_i }
                { \sqrt{{k_{n-1}}^+ {k_n}^+} }
           \Big]
           \Big[
           \frac{ \prod_{i=1}^{n} d{\bf k'}_i }
                { \sqrt{{k'_{n-1}}^+ {k'_n}^+} }
           \Big]
      \delta^{3}(\sum_{i=1}^{n}{\bf k}_i-\sum_{i=1}^{n}{\bf k'}_i)
\nonumber\\
      & &
      \hskip 45pt \times
      \prod_{i=1}^{n-2} \delta^{3}({\bf k}_i-{\bf k'}_i)
      \theta(\epsilon_{\Lambda_0}-
             \sum_{i=1}^{n}\epsilon_{{\bf k}_i})
      \theta(\epsilon_{\Lambda_0}-
             \sum_{i=1}^{n}\epsilon_{{\bf k'}_i})
\nonumber\\
      & &
      \hskip 45pt \times
      |{\bf k}_1,\dots,{\bf k}_n\rangle
      \langle{\bf k'}_1,\dots,{\bf k'}_n|,
\label{vn,n}
\nonumber\\
      v_{n,n+2}
      &=&
      \Big[
            v_{n+2,n}
      \Big]^{\dagger}
\nonumber\\
      &\equiv&
      \frac{\bar \lambda}{(n-1)!}
      \int
           \Big[
           \frac{ \prod_{i=1}^{n} d{\bf k}_i }
                { \sqrt{ {k_n}^+ } }
           \Big]
           \Big[
           \frac{ \prod_{i=1}^{n+2} d{\bf k'}_i }
                { \sqrt{ {{k'_n}^+} {k'_{n+1}}^+ {k'_{n+2}}^+ } }
           \Big]
      \delta^{3}(\sum_{i=1}^{n}{\bf k}_i-\sum_{i=1}^{n+2}{\bf k'}_i)
\nonumber\\
      & &
      \hskip 45pt \times
      \prod_{i=1}^{n-1} \delta^{3}({\bf k}_i-{\bf k'}_i)
      \theta(\epsilon_{\Lambda_0}-
             \sum_{i=1}^{n}\epsilon_{{\bf k}_i})
      \theta(\epsilon_{\Lambda_0}-
             \sum_{i=1}^{n+2}\epsilon_{{\bf k'}_i})
\nonumber\\
      & &
      \hskip 45pt \times
      |{\bf k}_1,\dots,{\bf k}_n\rangle
      \langle{\bf k'}_1,\dots,{\bf k'}_{n+2}|,
\label{vn,n+2}
\end{eqnarray}
where $\bar \lambda =\lambda/4!(2\pi)^3$, $\epsilon_{\bf k} \equiv
       ({{\bf k}_\perp}^2+\mu^2)/(2k^+)$ and $ \epsilon_{\Lambda_0}
       \equiv  ({{\bf P}_\perp}^2+\Lambda_0^2)/(2P^+) $.
The total energy and momentum of an $n$-body state
$
|{\bf k}_1,{\bf k}_2,\dots,{\bf k}_n \rangle
$
satisfies the dispersion relation
$
E=({{\bf P}_\perp}^2+M^2)/(2P^+),
$
so the invariant mass $M$ of the state is represented by the Jacobi
variables, $x_i$ and ${\bf r}_i$ ($i=1$ to $n$), as
\begin{equation}
              M^2= \sum_{i=1}^{n}
                 \frac{{{\bf r}_{i\perp}}^2+\mu^2}{x_i^+}
                 \geq {(n\mu)}^2,
\label{M}
\end{equation}
where
\begin{equation}
      {\bf k}_i \equiv (x_iP^+,x{\bf P}_{\perp}+{\bf r}_i).
\end{equation}
One of the Jacobi variables
is a dependent variable, since $\sum_{i=1}^{n}x_i=1$ and
$\sum_{i=1}^{n}{\bf r}_i=0$ in consequence of
${\bf P}=\sum_{i=1}^{n}{\bf k}_i$.
The $x_i$ can vary from 0 to 1 and two components of the vector
${\bf r}_i \equiv ({r_i}^{1},{r_i}^{2})$
from $-\infty$ to $\infty$.
The invariant mass $M$ becomes minimum $n\mu$ when ${\bf r}_i=0$
and $x_i=1/n$ for
all $i$, and it diverges at $r_i^j=\infty$ or at $x_i=0$.
The ultraviolet and infrared divergences are simultaneously
excluded by the invariant-mass regularization $M <\Lambda_0$.
The Hamiltonian (\ref{hlambda0}) has already been regularized
with this, that is, with
the step functions $\theta$ appearing in the momentum integrations.
Here, it should be noted that the Hamiltonian does not include
the zero mode at all. The Fock vacuum is obviously an eigenstate of
the Hamiltonian, indicating that the physical vacuum is trivial.
A center-of-mass motion is decoupled from internal motions in
the Hamiltonian, since it does not contain any interaction
between the two motions. This property is not broken by
the invariant-mass regularization.

The light-cone quantization mentioned does not treat the
zero mode properly\cite{my,ny}.
As shown in subsection \ref{rg23}, it is needless
for the symmetric phase, but not for the broken one.
In principle the zero mode is treat-able
by quantizing the theory in a spatial box $-L < x^- \le L$ under the
periodic boundary condition, but in practice the procedure
does not seem feasible, as mentioned in subsection \ref{rga}.
Ohio group suggests to add counterterms
to the regularized Hamiltonian which does not contain the zero mode
\cite {per1,pw,ohio}.
An aim of the present paper is to support the suggestion.

\subsection{The second and third steps of RG procedure}
\label{rg23}

The second step of our RG procedure starts with dividing
the finite transformation ($\Lambda_0 \rightarrow \Lambda$) into
successive small transformations
($\Lambda_0 \rightarrow \Lambda_1 \rightarrow \ldots \rightarrow
\Lambda$). For the $n$-th small transformation, $
G$ is obtained by solving
$
      G
      =
      G_{\rm 0}+
      G_{\rm 0}{\cal Q} V {\cal Q}G
$
perturbatively, where the projectors ${\cal P}$ and ${\cal Q}$ are
defined as
\begin{eqnarray}
          {\cal P}|{\bf k}_1,{\bf k}_2,\dots,{\bf k}_n\rangle
          &\equiv& \theta(\epsilon_{\Lambda_n}-\epsilon)
          |{\bf k}_1,{\bf k}_2,\dots,{\bf k}_n\rangle ,
\nonumber\\
          {\cal Q}|{\bf k}_1,{\bf k}_2,\dots,{\bf k}_n\rangle
          &\equiv& \theta(\epsilon_{\Lambda_{n-1}}-\epsilon)
                  \theta(\epsilon-\epsilon_{\Lambda_n})
          |{\bf k}_1,{\bf k}_2,\dots,{\bf k}_n\rangle,
\label{defp}
\end{eqnarray}
for any state having the total energy $\epsilon$.
As shown in (\ref{M}), $\Lambda_n$ should be larger than $\mu$,
so the one-body Fock state ( with $M=\mu$ )
can not be in the ${\cal Q}$ space. Any diagram including the one-body
state then does not contribute to the ${\cal Q}$-space Green function
$G$.

In this paper we make one-loop approximation and neglect all irrelevant
operators produced by the RG procedure as a reasonable approximation.
A perturbative expansion of the Bloch-Horowitz
Hamiltonian is then calculated up to second order
in $\lambda$, that is,
$
       H_{\Lambda}(M_i)
       = {\cal P}H_{0}{\cal P} + {\cal P}V{\cal P}
        +{\cal P}V{\cal Q}G_{0}{\cal Q}V{\cal P}
        +O(V^3).
$
The second-order correction ${\cal P}V{\cal Q}G_{0}{\cal Q}V{\cal P}$
generates two-, four- and six-point vertices, but the six-point one is
neglected, because it is an irrelevant operator. The correction is
obtained by calculating the matrix elements
$
\langle {\bf k}_1,{\bf k}_2, \dots,{\bf k}_n|
{\cal P}V{\cal Q}G_{0}{\cal Q}V{\cal P}
|{\bf k'}_1,{\bf k'}_2, \dots,{\bf k'}_m \rangle
$
 separately.

Matrix elements of
$
{\cal P}(V+V{\cal Q}G_{0}{\cal Q}V){\cal P}
$
between two-body Fock states are composed of
four-point vertices displayed in Fig. 2.
In our all diagrams, time flows toward the right.
The longitudinal and transverse momenta
are conserved at each vertex, and all particles are on shell.
A net contribution of the diagrams becomes
\begin{eqnarray}
      {\cal P} ( v_{22} + v_{22} G_0 v_{22} ) {\cal P}
       = \bar \lambda ( 1+9 \bar \lambda A)
         {\cal P} v_{22}{\cal P}.
\end{eqnarray}
The effective interaction has the same form as $V$,
but with a new coupling constant
$
  \bar \lambda(\Lambda_n)=\bar \lambda(\Lambda_{n-1})
                        \{ 1+9 \bar \lambda(\Lambda_{n-1}) A \},
$
where $A$ is a loop integral defined as
\begin{eqnarray}
 A &\equiv&
          \frac{1}{2}
          \int
          \Big[\prod_{i=1}^{4}
               \frac{d{\bf k}_i}{\sqrt{k_i^+}}
          \Big]
          \delta^{3}({\bf P} -
                       \sum_{i=1}^{2}{\bf k}_i)
          \langle {\bf k}_1,{\bf k}_2|G_0|{\bf k}_3,{\bf k}_4 \rangle
\nonumber\\
    &=&  \int
          \Big[\prod_{i=1}^{2}
               \frac{d{\bf k}_i}{k_i^+}
          \Big]
          \delta^{3}({\bf P} -
                       \sum_{i=1}^{2}{\bf k}_i)
          F({\bf k}_1,{\bf k}_2)
\nonumber\\
    &=& I_A(\Lambda_{n-1},\mu(\Lambda_{n-1}),M_i) -
        I_A(\Lambda_n,\mu(\Lambda_{n-1}),M_i)
\end{eqnarray}
for
\begin{equation}
F({\bf k}_1,{\bf k}_2,\cdots,{\bf k}_n)
          \equiv
          \frac{
               \theta(\epsilon_{\Lambda_{n-1}}-
                   \sum_{i=1}^{n}\epsilon_{{\bf k}_i})
               \theta(\sum_{i=1}^{n}\epsilon_{{\bf k}_i}-
                  {\epsilon_{\Lambda}}) }
               { E_i-\sum_{i=1}^{n}\epsilon_{{\bf k}_i} }.
\end{equation}
The loop integral $A$ is obtained with an analytic function $I_A$
defined in Appendix A. The function depends on the cutoff,
$\mu (\Lambda_{n-1})$ and $M_i$, but not on ${\bf P}$.
Hereafter, it is assumed that $\Lambda_0 \gg \Lambda_{phys}$ and
$M_i$ and $\mu(\Lambda_{n-1})$ are of order $\Lambda_{phys}$.
For $\Lambda_{n} \gg \Lambda_{phys}$, $A$ becomes $4 \pi t$, where
$
    t=\ln |\Lambda_{n}/\Lambda_{n-1}|.
$

Matrix elements of ${\cal P}(V+V{\cal Q}G_{0}{\cal Q}V){\cal P}$
between three-body states are also calculated in a similar way.
All diagrams contributing to the elements are also in Fig. 2, but
with a spectator added. An example is shown in Fig. 3.
The diagrams describe four-point vertices,
since one of three particles is a spectator.
A net contribution of these diagrams is
\begin{eqnarray}
   &&{\cal P} ( v_{33}+ v_{33} G_0 v_{33} ) {\cal P}
\nonumber\\
   &&=
      \frac{3}{2}
      \int
      \Big[
           \prod_{i=1}^{4}
           \frac{d{\bf k}_i}
                {\sqrt{{k_i}^+}}
      \Big]
      d{\bf p}
      \delta^{3}(\sum_{i=1}^{2}{\bf k}_i -
                   \sum_{i=3}^{4}{\bf k}_i)
      \theta(\epsilon_{\Lambda} -
             \sum_{i=1}^{2}\epsilon_{{\bf k}_i})
      \theta(\epsilon_{\Lambda} -
             \sum_{i=3}^{4}\epsilon_{{\bf k}_i})
\nonumber\\
      && \hskip 30pt \times
      \bar \lambda ( 1+ 9 \bar \lambda B({\bf p}))
      |{\bf k}_1,{\bf k}_2,{\bf p} \rangle
      \langle {\bf k}_3,{\bf k}_4,{\bf p}|,
\end{eqnarray}
where
\begin{eqnarray}
 B({\bf p}) &\equiv&
          \int
          \Big[\prod_{i=1}^{2}
               \frac{d{\bf k'}_i}{k_i^+}
          \Big]
          \delta^{3}({\bf P} -
                       \sum_{i=1}^{2}{\bf k'}_i -{\bf p})
          F({\bf k'}_1,{\bf k'}_2,{\bf p}).
\end{eqnarray}
The loop integral $B({\bf p})$ depends on ${\bf p}$,
a momentum of the spectator. As shown in Fig. 4 and Appendix A,
$B({\bf p})$ little differs from $A$ for $\Lambda_{n} \gg
\Lambda_{phys}$,
but the difference becomes significant as $\Lambda_{n}$ decreases;
it is appreciable
at $\Lambda_{n} \sim 10\Lambda_{phys}$ and sizable at
$\Lambda_{n} \sim 5\Lambda_{phys}$. This allows $H_{\Lambda_{n}}$
to depend on the momentum of the spectator and the number of particles
in the initial and final states.
The dependences stem from the fact that
the invariant mass regularization adopted breaks
covariance and cluster decomposition. In this paper
$B({\bf p})$ is set to $A$ as a reasonable approximation, and
$\Lambda_{n}$ is then considered to be larger than $10\Lambda_{phys}$.
Only an exception is subsection \ref{effpot} in which
the effective potential is calculated from
a solution at the small $\Lambda$ limit to RG equations.
In this sense the calculation contains errors to some extent.
A way of treating the
spectator dependence explicitly is proposed in \cite{per1}. It may be
useful for the present RG scheme, when one has to treat the dependence
accurately.

Other matrix elements of
${\cal P}( V+V{\cal Q}G_{0}{\cal Q}V ) {\cal P}$
are also derivable in a similar fashion:
\begin{eqnarray}
    {\cal P}( v_{31}+ v_{33} G_0 v_{31}) {\cal P}
      &=&
      \Big[
       {\cal P} ( v_{13} + v_{13} G_0 v_{33} ) {\cal P}
      \Big]^\dagger
\nonumber\\
      &=&
      \int
          \Big[
               \prod_{i=1}^{4}
               \frac{d{\bf k}_i}
                    {\sqrt{{k_i}^+}}
          \Big]
      \theta(\epsilon_{\Lambda_1}-
             \sum_{i=1}^{3}\epsilon_{{\bf k}_i})
      \theta(\epsilon_{\Lambda_1}-\epsilon_{{\bf k}_4})
      \delta^{3}(\sum_{i=1}^{3}{\bf k}_i-{\bf k}_4)
\nonumber\\
      & & \hskip 3pt \times
      \bar \lambda(\Lambda_{n-1})
      \{ 1+9 \bar \lambda(\Lambda_{n-1}) B({\bf k}_3) \}
      |{\bf k}_1,{\bf k}_2,{\bf k}_3\rangle
      \langle{\bf k}_4|
\nonumber\\
      &\approx&
      \bar \lambda(\Lambda_{n-1})
      \{ 1+9 \bar \lambda(\Lambda_{n-1}) A \}
      {\cal P}v_{33}{\cal P}.
\end{eqnarray}
Figure 5 shows a unique one-loop diagram contributing
to the matrix element ${\cal P}( v_{31}+ v_{33} G_0 v_{31}) {\cal P}$.
No one-loop diagram contributes to the matrix element
${\cal P} v_{13} G_0 v_{31} {\cal P}$, indicating that
the element vanishes within the present approximation.

The effective Hamiltonian $H_{\Lambda_{n}}$ thus obtained holds
the same form as $H_{\Lambda_{n-1}}$, but with new parameters
\begin{eqnarray}
      \bar \lambda(\Lambda_{n})
      =
      \bar \lambda(\Lambda_{n-1})
      \{ 1+ 9 \bar \lambda(\Lambda_{n-1}) A \},
\hskip 20pt
      \mu(\Lambda)^2
      = {\mu(\Lambda_{n-1})}^2.
\label{rgeq1}
\end{eqnarray}
Taking the limit $\Lambda_{n} \rightarrow \Lambda_{n-1}$ leads to
RG equations
\begin{eqnarray}
      \frac{d\lambda}{dt}
      =  \frac{3 \zeta}{4\pi} \frac{dI_A(\Lambda,\mu(\Lambda))}{dt}
         \lambda^2 +O(\hbar^2),
\hskip 20pt
      \frac{d\mu^2}{dt}
      &=& 0 + O(\hbar^2),
\label{GL0}
\end{eqnarray}
where $\zeta=\hbar/(16\pi^2)$ and $t=\ln(\Lambda/\Lambda_0)$.
For $\Lambda \gg \Lambda_{phys}$,
the right hand side of the first equation
tends to $ 3 \zeta \lambda^2 $. The equations are consistent
with those calculated in the covariant perturbation theory,
as shown in subsection \ref{effpot}.
The right hand side originally included a partial derivative
$  \partial I_A(\Lambda,\mu(\Lambda)) / \partial t $
with $\mu(\Lambda)$ fixed, but it has been replaced by $dI_A/dt$.
The replacement is correct in this case, because of $d\mu/dt=0$.
Even if $d\mu/dt \ne 0$ just as in the case of subsection \ref{rga},
one
can make the same replacement
to derive RG equations at one-loop level,
since $dI_A/dt=\partial I_A/\partial t+O(\hbar)$ as a result of
$d\mu/dt=O(\hbar)$. The solution to (\ref{GL0}) is
\begin{eqnarray}
      \lambda(\Lambda)
      =
      \frac{\lambda_0}{1-3 \zeta \tilde A \lambda_0 /(4\pi)},
\hskip 20pt
      \mu(\Lambda)^2
      = \mu_0^2,
\label{solsym}
\end{eqnarray}
where $\lambda_0=\lambda(\Lambda_0)$, $\mu_0=\mu(\Lambda_0)$ and
$\tilde A=I_A(\Lambda_0,\mu(\Lambda_0))- I_A(\Lambda,\mu(\Lambda))$.
Expanding the solution $\lambda(\Lambda)$ in power of $A$, one can
find that it contains not only
contributions of the one-loop diagrams (Fig. 2)
but also those of the ladder
diagrams, although for each small transformation
only the one-loop diagrams have been taken into account.

The RG procedure ends up with the scale transformation $T$ of
(\ref{scale:p}) and (\ref{scale:h}), where it is assumed that
there exists only a Gaussian fixed point in the present model.
The transformed Hamiltonian
$
   H_{\Lambda_{0}}' \equiv T[H_{\Lambda}]
$
has parameters $\lambda(\Lambda)'=\lambda(\Lambda)$ and
      $\mu(\Lambda)'^2 = (\Lambda_0/\Lambda)^2 \mu(\Lambda)^2$.
The running coupling $\lambda(\Lambda)'$
depends not only on $\Lambda$,
but also on the eigenvalue $M_i$ of $H_{\Lambda_0}$
through $\tilde A$.
In the limit $\Lambda=\infty$, however, the coupling tends to
$\lambda_0/(1-3 \zeta \lambda_0 t)$, indicating no $M_i$ dependence.
The mass parameter $\mu(\Lambda)$ present in $H_{\Lambda}$
has no $\Lambda$ dependence and then
equal to the lowest mass $M_1$, so that the corresponding parameter
$\mu(\Lambda)'$ in the transformed Hamiltonian $T[H_{\Lambda}]$
behaves as $M_1 \Lambda_0/ \Lambda$.
We then draw a flow diagram for the lowest
state by setting $M_i=M_1=M_{phys}$ in $\tilde A$ as a renormalization
condition. The diagram presented in Fig. 6 shows, as expected,
that there exists a
renormalization trajectory on the positive $\mu'^2$ axis.
On the axis $\lambda'$ is always zero, indicating
that the present model is trivial. Once the trajectory is found
for the lowest state, renormalization
trajectories for other states
are obtainable from the one for the lowest
state by replacing $M_1$ by invariant masses of the excited states,
if they are known.
In the present case, such a state-dependence does not appear
as a reflection of the triviality, that is, all the trajectories
exist on the positive $\mu'^2$ axis,

There appear two phases in Figure 6; a critical line between
the two is on the positive $\lambda'$ axis.
A phase present
in the first quadrant ($\lambda' \ge 0$ and $\mu'^2 \ge 0$) is
symmetric for the reflection $\phi \rightarrow -\phi$, since
the effective Hamiltonian holds the symmetry.
Another phase appearing in the second quadrant is then
considered to be
a broken one. The present cutoff Hamiltonian $H_{\Lambda_0}$, however,
is not applicable for the phase,
since Fock states with negative $\mu^2$ are not physical
in the sense that tachyons come out.
In general, light-front Hamiltonians are different
between the phases,
while their vacua are always equal to the Fock vacuum.
This is explicitly shown
in two-dimensional $\phi^4$ model \cite{ps,bps,bsh};
for the symmetric phase
the Hamiltonian contains oscillatory modes only,
while for the broken phase
it includes both zero and oscillatory ones.
The Hamiltonian for the broken phase has a new mass term produced
by the zero mode in addition to
the original one $\mu^2$, so one can define Fock states with the sum,
even if $\mu^2$ is negative. Further discussion on the broken phase is
made in sec. \ref{rga}.

\subsection{The RG equations for the broken phase}
\label{rga}
The RG equation (\ref{GL0}) is valid for the symmetric phase, but not
for the broken one, as shown in subsection \ref{rg23}. The
failure stems from the fact that the present $H_{\Lambda_0}$
does not contain the zero ($k^+=0$) mode.
The zero mode is responsible for the phase transition, because
the order parameter $\langle 0 |\phi|0 \rangle$ can not become nonzero
without the zero mode, that is, the order parameter is
zero for any oscillator ($ k^+ > 0 $) mode.
Hence the present $H_{\Lambda_0}$ is valid
only for the symmetric phase
in which $\langle 0 |\phi|0 \rangle =0 $.

There are two ways of finding a Hamiltonian
valid for the broken phase. One way is to treat the zero mode
explicitly\cite{my,ps,bps,bsh}.
For this purpose $\phi^4$ theory is usually quantized
in a spatial box $-L < x^- \le L$
under the periodic boundary condition,
so that the zero mode is
separated from the oscillator ( $k^+ > 0$ ) ones.
The resulting bare Hamiltonian is
different from the present $H_{\Lambda_0}$
in the sense that the bare Hamiltonian contains
not only the oscillator modes but also the zero mode.
The zero mode is an operator-valued function of
the oscillator modes, since the zero mode satisfies a
constraint \cite{my}
\begin{eqnarray}
\int_{-L}^{L} dx^- \left\{
                  \left( \mu^2-\partial_{\perp}^2 \right)\phi(x)
                  + \frac{\lambda}{3!}\phi^3(x)
                  \right\} = 0.
\label{zero}
\end{eqnarray}
This has been obtained by integrating the equation of motion
over $x^-$. In general, it is not easy to solve the operator-valued
nonlinear equation for the zero mode without perturbation
in order to treat spontaneous symmetry breaking.
A trial is reported for $\phi^4$
model in two dimensions. In the work \cite{ps}, the equation
is symmetrically ordered and approximately solved
under the Tamm-Dancoff truncation.
However, it is not clear how we should
order the equation, since the zero mode is not a dynamical operator.

Another way is to add counterterms
to the present $H_{\Lambda_0}$ which
has no zero mode \cite{per1,pw,ohio}.
Only the zero mode can make the order parameter
$\langle 0 |\phi|0 \rangle$ nonzero.
This means that the system has to sit in the bottom of the effective
potential, as soon as the zero mode is discarded from $\phi$.
So we shift $\phi$ as $\phi \rightarrow \phi-v$, and determine
a value of $v$ on a renormalization trajectory.
In terms of the shifted field $\phi$ the Lagrangian density becomes
\begin{eqnarray}
    {\cal L} &=& \frac{1}{2}\partial_{\mu}\phi\partial^{\mu}\phi
                -\frac{1}{2}m^2\phi^2
                -w \phi
                -\frac{g}{3!}\phi^3
                -\frac{\lambda}{4!}\phi^4,
\end{eqnarray}
where
\begin{eqnarray}
       w=v\mu^2 + \frac{\lambda}{3!}v^3,
\hskip 10pt
       m^2 = \mu^2 + \frac{\lambda v^2}{2},
\hskip 10pt
       g=v \lambda,
\label{newpara}
\end{eqnarray}
where a constant term has been discarded,
because it is irrelevant to physics. The light-front
Hamiltonian is then
\begin{equation}
    H = \int d^3 x
           \bigg \{
                 \frac{1}{2}\Big[
                 \sum_{i=1}^{2}(\partial_i \phi)^2+
                   m^2 \phi^2 \Big]
                 + \frac{g}{3!}:{\phi^3}:
                 + \frac{\lambda}{4!}:{\phi^4}: \bigg \}.
\end{equation}
As a feature of the naive light-front field theory which
does not treat the zero mode, the linear term in $\phi$ vanishes.
Following the RG procedure shown in the previous subsections, one has
RG equations for $m^2$, $g$ and $\lambda$.
Inserting (\ref{newpara}) into the equations yields
RG equations for the original parameters $\mu^2$, $\lambda$ and
$v$,
\begin{eqnarray}
      \frac{d\lambda}{dt}
      =
      \frac{3 \zeta}{4\pi} \frac{dI_A}{dt} \lambda^2,
\hskip 20pt
      \frac{dv}{dt}
      = 0,
\hskip 20pt
      \frac{d\mu^2}{dt}
      = - \frac{\zeta}{8\pi} \frac{dI_A}{dt} (\lambda v)^2.
\label{GLOP}
\end{eqnarray}
The solution to (\ref{GLOP}) is easily obtained as
\begin{eqnarray}
      \lambda(\Lambda)
      &=&
      \frac{\lambda_0}{1-3 \zeta \tilde A \lambda_0/(4\pi)},
\nonumber\\
      v(\Lambda)
      &=&
      v_0,
\nonumber\\
      \mu(\Lambda)^2
      &=&
          \mu_0^2 +
          \frac{\lambda_0 v_0^2}{6}
          \left \{1- \frac{1}{1-3\zeta \tilde A \lambda_0/(4\pi)}
          \right \},
\label{sol1}
\end{eqnarray}
where all parameters at $\Lambda=\Lambda_0$ are characterized
by a subscript ``0''.
For $\Lambda \gg \Lambda_{phys}$,
$A$ tends to $4\pi t$ in (\ref{sol1}),
indicating no $M_i$ dependence of the solution in the limit.
The corresponding parameters, $\mu'$, $\lambda'$ and
$v'$, present in $T[H_{\Lambda}]$ are then obtained by
$\mu'(\Lambda)=\mu(\Lambda)\Lambda_0/\Lambda$,
$\lambda'(\Lambda)=\lambda(\Lambda)$ and
$v'(\Lambda)=v(\Lambda)\Lambda_0/\Lambda$.
Obviously, one can find a renormalization trajectory on the positive
$v'^2$ axis in the parameter space $(\lambda',\mu'^2,v'^2)$,
as shown in Fig. 7. The trajectory does not depend on $M_i$,
since $\lambda(\Lambda)'$ and $\mu(\Lambda)'$ are always zero
on the axis. The trajectory has thus no state dependence
in the present case.

In (\ref{GLOP}), the sum of $v^2/6$ times the first equation
and the third one becomes
$d(\mu^2 + \lambda v^2/6)/dt=0$, indicating that
$C \equiv \mu^2 + \lambda v^2/6$ is RG invariant.
Thus, $v$ is a dependent variable obtained by
$ \lambda v^2/6 = - \mu^2 + C$. This is a natural result of the fact
that the original theory includes only $\mu^2$ and
$\lambda$ as physical parameters.
The relation between $v^2$ and the physical parameters
should be unique. In fact, $C$ is determined as follows.
The relation becomes $ \lambda' v'^2/6 = - \mu'^2 + C \exp(2t)$
for parameters $\lambda'$, $v'^2$ and $\mu'^2$,
and it forms a group of
surfaces, each with different $C$, in the parameter space.
Only a surface with $C=0$ contains the critical
line ( the positive $\lambda'$ axis ) between the broken and symmetric
phases. The line is a border of the surface,
since $\mu'^2 = -\lambda' v'^2/6 \le 0$ for positive $\lambda'$.
The surface can be regarded as the broken phase,
since it can be connected to the symmetric one on the critical line.
The effective potential should become minimum on the surface, and
it is really confirmed in Sec. \ref {effpot}. The surface is
also depicted in Fig. 7. The renormalization trajectory
is on the surface, as expected.
Perry and Wilson also find the same relation
without calculating the effective
potential\cite{pw}. Their derivation is essentially equal to ours,
although $C$ is set to zero a priori in their derivation.

\subsection{Covariant perturbation theory and effective potential}
\label{effpot}

The RG equations (\ref {GLOP}) based on the present light-front
Hamiltonian formalism are compared with those on
the covariant (equal-time) perturbation theory. For this purpose
we use a cutoff on the Euclidean momentum, $\Lambda^2 < k^2$,
and make the same approximations,
that is, the one-loop approximation and
dropping irrelevant operators.
The resultant RG equations are
\begin{eqnarray}
      \frac{d\lambda}{dt}
      &=&
       \frac{3 \zeta \lambda^2 }{(1+r)^2} ,
\hskip 20pt
      \frac{dg}{dt}
      =
      \frac{ 3 \zeta \lambda g}{(1+r)^2},
\hskip 20pt
      \frac{dm^2}{dt}
      =
       - \frac{\zeta\lambda\Lambda^2}{1+r}
       + \frac{\zeta g^2}{(1+r)^2},
\nonumber\\
      \frac{dw}{dt}
      &=&
       - \frac{\zeta g \Lambda^2}{1+r},
\label{GLC}
\end{eqnarray}
where $r=m^2/\Lambda^2$.
Inserting relation (\ref{newpara})
 into (\ref{GLC}) yields RG equations
for the original parameters,
\begin{eqnarray}
      \frac{d\lambda}{dt}
      &=&
       \frac{3 \zeta \lambda^2 }{(1+r)^2} ,
\hskip 20pt
      \frac{dv}{dt}
      =
      0 ,
\hskip 20pt
      \frac{d\mu^2}{dt}
      =
       - \frac{\zeta\lambda\Lambda^2}{1+r}
       + \frac{\zeta (\lambda v)^2}{2(1+r)^2}.
\label{GLCO}
\end{eqnarray}
In the derivation of (\ref{GLCO}) from (\ref{GLC}), the number of
equations is reduced from 4 to 3. One of four equations
in (\ref{GLC}) is thus not independent under
the condition (\ref{newpara}).
For $\Lambda$ much larger than $m$ and $M_i$, (\ref{GLCO}) agree with
(\ref{GLOP}), except
(\ref{GLCO}) newly have a term $-\zeta \lambda \Lambda^2/(1+r)$
in its third equation.
The term comes from the tadpole (Fig.1), but in the present
formalism the self mass has already been included
in the mass parameter $\mu$ by taking the normal ordered Hamiltonian.
Hence, the RG equations obtained in the light-front Hamiltonian
formalism are essentially equal to equations (\ref{GLCO}) calculated
in the covariant perturbation theory.
Of course, the two RG equations are not identical except the large
$\Lambda$ limit, because regularization schemes
are different between the two formulations.
The light-front perturbation theory is formally
 equivalent to the covariant
one \cite{chang}, if the same regularization scheme is taken.
In fact, as soon as the $k^-$ integration is made
in the covariant formalism, one finds a direct connection between
diagrams obtained in the covariant theory and time-ordered ones
in the present formalism.

The equality guarantees that within the framework of the light-front
Hamiltonian formalism we can calculate the effective potential
through the following ordinary procedure. In the covariant perturbation
theory, the effective potential
$V(\phi_{cl})$ is obtained as
$\mu_R^2 \phi_{cl}^2/2 + \lambda_R \phi_{cl}^4/4!$
within the present approximations, where $\mu_R$ and $\lambda_R$ stand
for renormalized quantities and  $\phi_{cl}$ for the classical field,
that is,
the vacuum expectation value ($v$) of $\phi$ which is not shifted.
The renormalized quantities are solutions $\mu(\Lambda)$ and
$\lambda(\Lambda)$
at the small $\Lambda$ limit to RG equations (\ref{GLCO})
under the condition $v=0$, because $\mu_R$ ($\lambda_R$)
are just the sum of one-particle irreducible diagrams produced by an
interaction $\lambda\phi^4$ with 2 (4) external lines.
In fact, the solutions are
\begin{eqnarray}
  \Bigl.
  \lambda(\Lambda)
  \Bigr |_{\Lambda=0}
  &=& \lambda_0
  -  \frac{3}{2} \zeta \lambda_0^2
  \left \{
    \ln \left |  \frac{\Lambda_0^2+\mu_0^2}{\mu_0^2} \right |
    -      \frac{\Lambda_0^2}{\Lambda_0^2+\mu_0^2}
  \right \} + O(\hbar^2)
  \nonumber\\
  \Bigl.
  \mu(\Lambda)^2
  \Bigr |_{\Lambda=0}
  &=&
  \mu_0^2
  -  \frac{1}{2} \zeta \lambda_0
  \left \{
    \Lambda_0^2 -
    \mu_0^2 \ln \left |  \frac{\Lambda_0^2+\mu_0^2}{\mu_0^2} \right |
  \right \} + O(\hbar^2)
\label{solcov}
\end{eqnarray}
at the one-loop level, and they agree with $\mu_R$ and $\lambda_R$
directly calculated with the covariant perturbation theory.
Also in the light-front Hamiltonian formalism,
$\mu_R$ and $\lambda_R$ are obtained from
the solutions (\ref{sol1}) by setting
$\Lambda$ to the small $\Lambda$ limit ($2\mu$) and $v$ to zero.
The $\lambda_R$ thus obtained contains not only
contributions of the one-loop diagrams (Fig. 2) but also those of the
ladder diagrams. All the contributions should be included
in $\lambda_R$, since they all are one-particle irreducible diagrams.
It is then found that $\mu_R=\mu_0$ and
$\lambda_R=\lambda_0/(1-a\lambda_0)$
for $a=3\zeta \tilde A/(4\pi)$ at $\Lambda=2\mu$:
Conversely, (i) $\mu_0=\mu_R$ and
$\lambda_0=\lambda_R/(1+a\lambda_R)$.
The condition $dV(\phi_{cl})/d\phi_{cl}=0$ on which
the effective potential is minimum leads to
(ii) $6\mu_R^2= - \lambda_R \phi_{cl}^2$.
RG equations (\ref{GLOP}) are now solved
under the initial conditions (i) and (ii); of course
$\phi_{cl}$ is identified with $v$.
As mentioned in sec. \ref{rga},
$C \equiv \mu(\Lambda)^2 + \lambda(\Lambda) v(\Lambda)^2/6$
is RG invariant. For any $\Lambda$ it becomes
$-a v_0^2\lambda_R^2/(1+a\lambda_R)$,
because of relations (i) and (ii).
In the large $\Lambda_0$ limit, the quantity $a$
diverges, so that $C$ tends to $-v_0^2\lambda_R$ for
$\lambda_R=\lambda_0/(1-a\lambda_0) \rightarrow 0$.
Hence, $C$ is zero in the limit, as far as $v_0$ is finite.
This is a reflection of the fact that
$\lambda_R$ is forced to vanish in the limit, that is,
the present model is trivial.  The condition for the system to be in
the bottom of the effective potential is then
$\mu(\Lambda)^2 + \lambda(\Lambda) v(\Lambda)^2 /6=0$. It agrees
with the result shown in subsection \ref{rga}.

\acknowledgments
We would like to acknowledge stimulating discussions with our
colleagues in Kyushu University, in particular S. Tominaga.
Support of this work is provided by a Grant-in-Aid for Scientific
Research from the Ministry of Education, Science and Culture of Japan
(No. 06640404).

\appendix
\section{Loop integrals}
The loop integral $A$ is easily performed by introducing
the Jacobi coordinate $(x,{\bf r})$ defined with
$
      {\bf k}_1 = (xP^+, x{\bf P}_\perp + {\bf r}),
      {\bf k}_2 = ((1-x)P^+, (1-x){\bf P}_\perp - {\bf r}).
$
The result is $ A = I_A(\Lambda_{n-1})-I_A(\Lambda_n) $ with
\begin{eqnarray}
I_A(\Lambda)
      &\equiv&
      2 \int dx d{\bf r}
      \frac{\theta(x(1-x)\Lambda^2 - {\bf r}^2 - \mu^2)}
           {x(1-x){M}^2 - {\bf r}^2 - \mu^2}
\nonumber\\
      &=&
      2\pi \ln \left |
          \frac{ 1-\sqrt{a(\Lambda)} }{ 1+\sqrt{a(\Lambda)} }
         \right |
      - 2\pi \sqrt{ a(M) }
      \ln \left |
          \frac{ \sqrt{ a(\Lambda) }-\sqrt{ a(M) } }
               { \sqrt{ a(\Lambda) }+\sqrt{ a(M) } }
         \right |
      \theta(a(M))
\nonumber\\
      & & {} +
      4\pi \sqrt{- a(M) }
      \arctan \left \{
          \frac{\sqrt{a(\Lambda)}}
               {\sqrt{a(-M)}}
              \right \}
      \theta(-a(M))
\end{eqnarray}
for $\Lambda \ge 2\mu$ and zero for $\Lambda < 2\mu $,
where $a(z)=1-4\mu^2/z^2$.

The loop integral $B$ is also given in a similar fashion.
As an example let us consider the diagram of
Fig. 3, in which particles 1 and 2 interact with each other, while
particle 3 is free. Each has a momentum
${\bf k}_i $ in the initial state and ${\bf k}'_i $ in the
intermediate state. For convenience we introduce
the Jacobi coordinates
\begin{eqnarray}
      {\bf k}_1 &=& (x(1-y)P^+,
       x[(1-y){\bf P}_\perp-{\bf s}] + {\bf r}),
\nonumber\\
      {\bf k}_2 &=& ((1-x)(1-y)P^+,
      (1-x)[(1-y){\bf P}_\perp-{\bf s}] - {\bf r}),
\nonumber\\
      {\bf k}_3 &=& (yP^+, y{\bf P}_\perp + {\bf s}).
\end{eqnarray}
The momentum (x,{\bf r}) represents a relative motion
between interacting two particles in the initial state,
and (x',{\bf r}') the motion in the intermediate state.
On the other hand, (y,{\bf s}) is a conserved momentum describing
a relative motion between the interacting pair and
the third particle.
The quantity $B$ is
obtained by making integrations over $x'$ and ${\bf r}'$,
so it is eventually given as a function of $y$ and ${\bf s}$:
$B \equiv I_B(\Lambda_{n-1})-I_B(\Lambda_n)$ with
\begin{eqnarray}
I_B(\Lambda)
      &=&
      2\pi
      \ln \left |
          \frac{1-\sqrt{b(\Lambda)}}{1+\sqrt{b(\Lambda)}}
         \right |
      - 2\pi
      \sqrt{b(M)}
      \ln \left |
          \frac{\sqrt{b(\Lambda)}-\sqrt{b(M)}}
               {\sqrt{b(\Lambda)}+\sqrt{b(M)}}
         \right |
      \theta(b(M))
\nonumber\\
      & & {} +
      4\pi \sqrt{-b(M)}
      \arctan
         \left \{
            \frac{\sqrt{b(\Lambda)}}
                 {\sqrt{-b(M)}}
         \right \}
      \theta(-b(M))
\end{eqnarray}
for $\Lambda \ge {\cal M}$ and zero for $\Lambda < {\cal M} $,
where $s^2 \equiv s_1^2+s_2^2$ and
\begin{equation}
  b(z)
  \equiv
  \frac{(z^2-{\cal M}^2)(1-y)}{(z^2-{\cal M}^2)(1-y)+(2\mu)^2},\quad
  {\cal M}^2
  \equiv
  \frac{(2\mu)^2}{1-y} + \frac{\mu^2}{y} + \frac{s^2}{y(1-y)}
  \ge (3\mu)^2 .
\end{equation}
The invariant mass $M$ of the
initial state is smaller than $\Lambda_{n-1}$, since the state is
in the ${\cal P}$ space. The mass is related to ${\cal M}$ as
\begin{eqnarray}
      M^2
           \equiv
             \frac{\mu^2+r^2}{(1-y)x(1-x)}
            +\frac{\mu^2}{y}
            +\frac{s^2}{y(1-y)}
           \ge {\cal M}^2 \ge (3\mu)^2,
\label{m0ini}
\end{eqnarray}
where use has been made of the inequality
$(\mu^2+r^2)/(x(1-x)) \ge (2\mu)^2$. Equation (\ref{m0ini})
indicates that ${\cal M}$ is somewhere between $3\mu$ and $M$.
The ${\cal M}$ is conserved during the process Fig. 3,
since it
is a function of the conserved momentum (y,{\bf s}).
In Fig. 4,
$B$ is drawn as a function of ${\cal M}^2$ and $y$
(instead of $s^2$ and $y$) and compared with $A$.
For $\Lambda_{n-1} =100\Lambda_{phys}$, in Fig. 4(a),
$B$ agrees with $A$ at all ${\cal M}$ and $y$.
The agreement becomes poor gradually
as $\Lambda$ decreases to $2\mu$.  For example,
for $\Lambda_{n-1} = 10 \Lambda_{phys}$ in Fig. 4(b),
$B$ is close to $A$ at ${\cal M} < 0.8 \Lambda_{n-1}$,
but undershoots $A$
for $0.9\Lambda_{n-1} < {\cal M} < \Lambda_{n-1}$, indicating that
$B$ can be approximated into $A$
for any initial state with $M$ less than
$0.8\Lambda_{n-1}$.
%
%
%
%
%
%
%

\begin{figure}[h]
\begin{center}
\epsfile{file=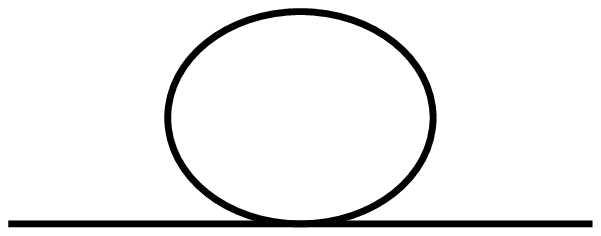,scale=0.8}
\end{center}
\caption{
Tadpole(one-loop self-mass) diagram.
The divergent contribution of the diagram is renormalized
by a redefinition of the mass parameter $\mu^2$.
}
\label{fig1}
\end{figure}
\begin{figure}[h]
\begin{center}
\epsfile{file=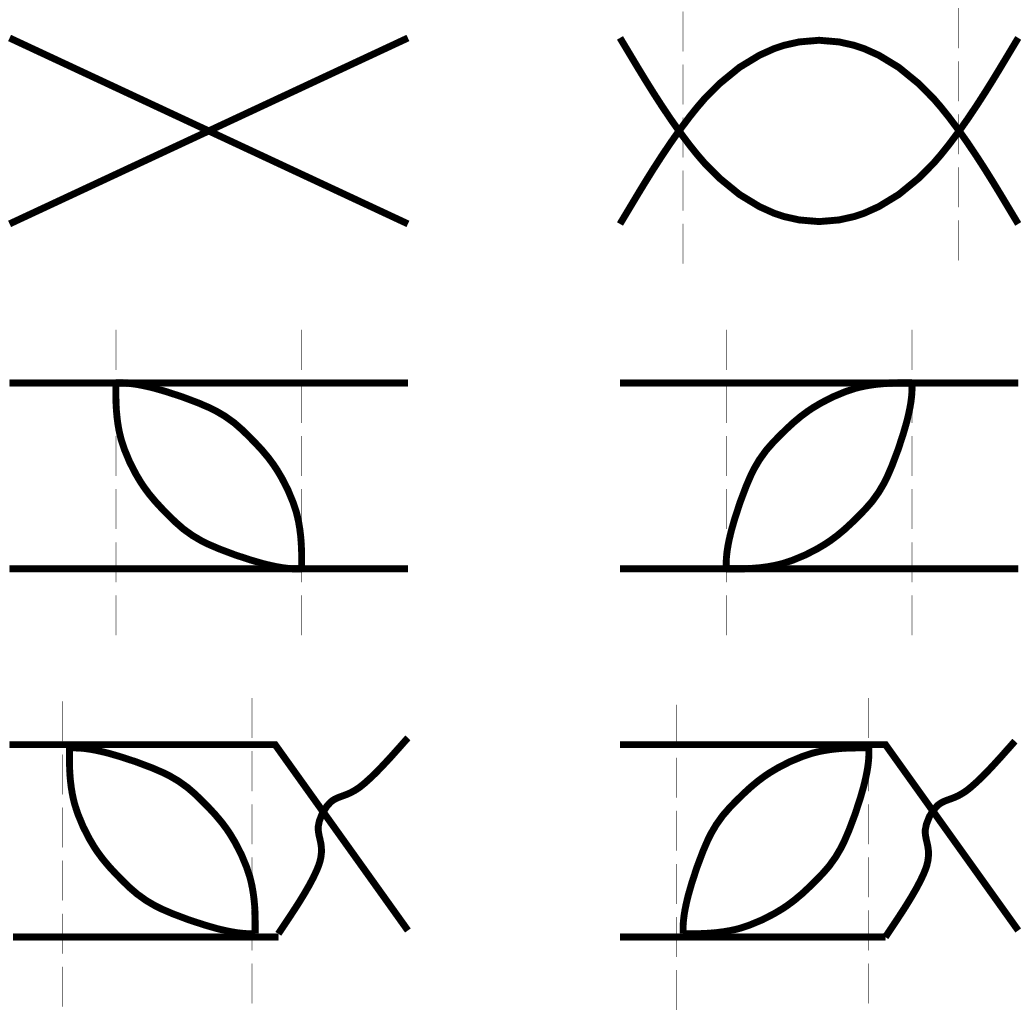,scale=0.8}
\put(-95,175){${\cal P}$}
\put(-54,175){${\cal Q}$}
\put(-10,175){${\cal P}$}
\put(-230,95){${\cal P}$}
\put(-195,95){${\cal Q}$}
\put(-155,95){${\cal P}$}
\put(-90,95){${\cal P}$}
\put(-53,95){${\cal Q}$}
\put(-15,95){${\cal P}$}
\put(-235,10){${\cal P}$}
\put(-205,10){${\cal Q}$}
\put(-165,10){${\cal P}$}
\put(-95,10){${\cal P}$}
\put(-61,10){${\cal Q}$}
\put(-20,10){${\cal P}$}
\end{center}
\caption{
One-loop diagrams
which contribute to the matrix elements of
${\cal P}(V+V{\cal Q}G_{0}{\cal Q}V){\cal P}$
between two-body states.
}
\label{fig2}
\end{figure}
\begin{figure}[h]
\begin{center}
\epsfile{file=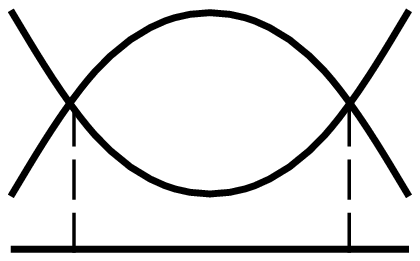,scale=0.8}
\put(-95,-10){${\cal P}$}
\put(-54,-10){${\cal Q}$}
\put(-10,-10){${\cal P}$}
\end{center}
\caption{
An example of the one-loop diagrams
which contribute to the matrix elements of
${\cal P}(V+V{\cal Q}G_{0}{\cal Q}V){\cal P}$
between three-body states.
}
\label{fig3}
\end{figure}

\begin{figure}[h]
\begin{center}
\epsfile{file=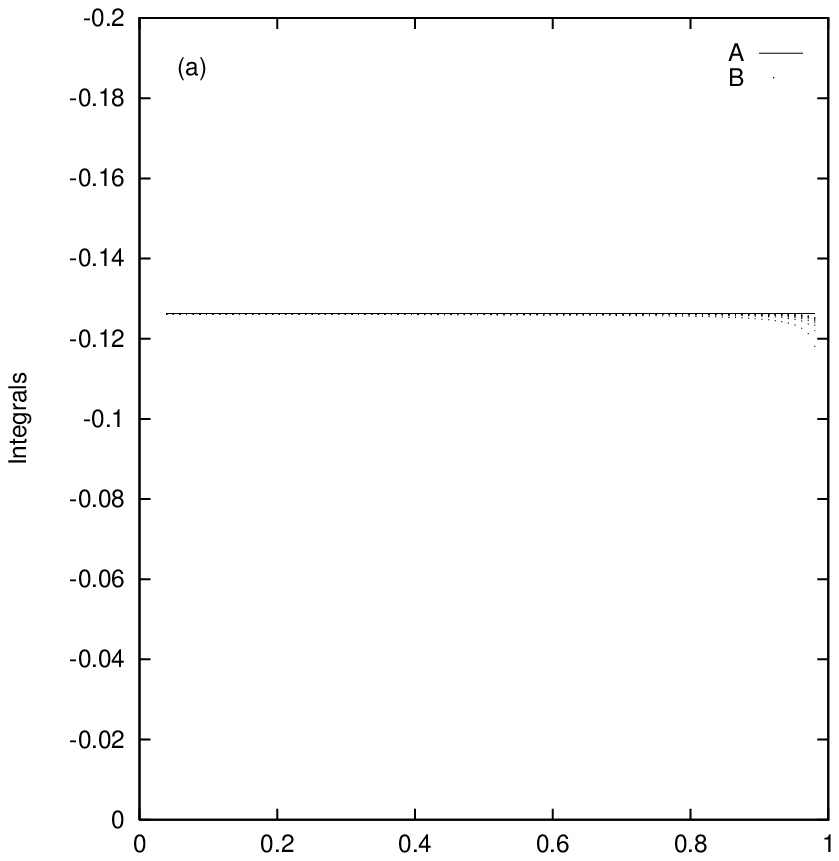,scale=0.9}
\epsfile{file=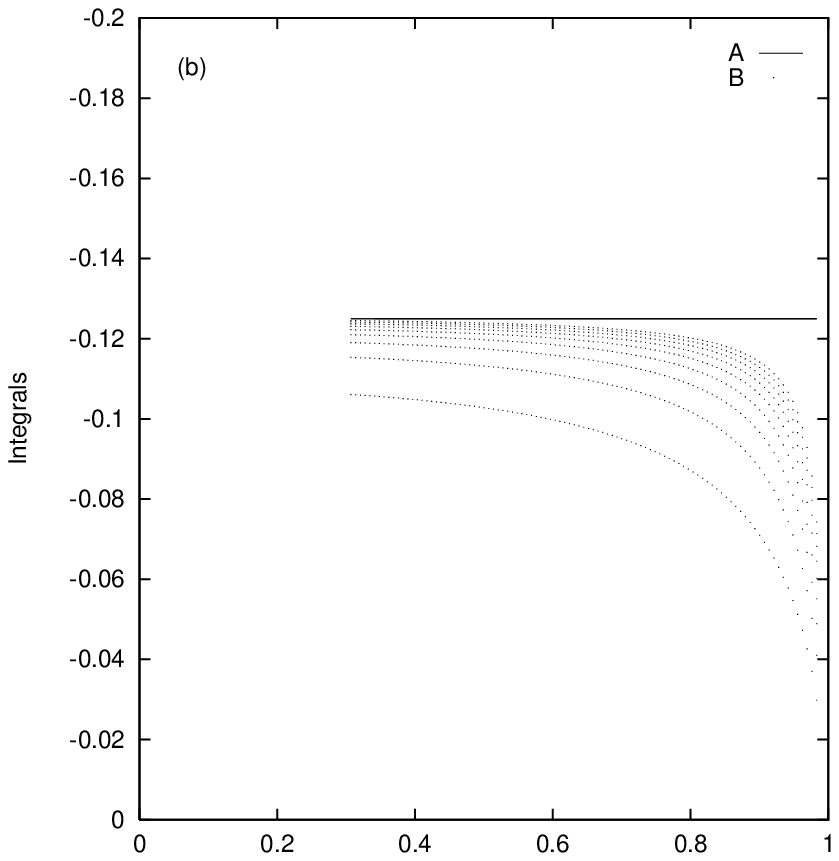,scale=0.9}
\vspace{1.0cm}
\put(-360,-15){${\cal M}/\Lambda_{n-1}$}
\put(-125,-15){${\cal M}/\Lambda_{n-1}$}
\end{center}
\caption{ The loop integrals $A$ and $B$
as a function of ${\cal M}$ and $y$;
these are defined in Appendix A.
The integral $A$ is plotted by the solid line, and
$B$ by dots. Each curve, which is given by connecting the dots,
has different $y$,
from $0.1$ to $0.9$ with an interval $0.1$:
$B$ decreases as $y$ goes down with ${\cal M}$ fixed.
Different parameter sets are taken in (a) and (b):
(a)$\Lambda_{\rm phys}=M_i=\mu=0.01\Lambda_{n-1}$.
(b)$\Lambda_{\rm phys}=M_i=\mu=0.1\Lambda_{n-1}$.
}
\label{fig4}

\begin{figure}[h]
\begin{center}
\epsfile{file=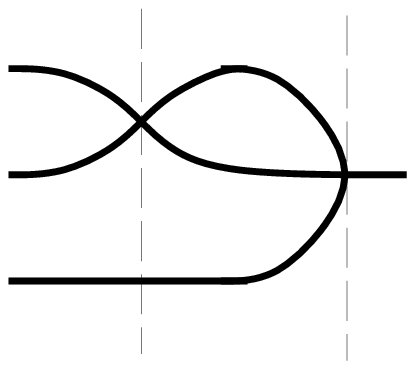,scale=0.8}
\put(-85,-10){${\cal P}$}
\put(-45,-10){${\cal Q}$}
\put(0,-10){${\cal P}$}
\end{center}
\caption{
A unique one-loop diagram which contributes to the matrix elements
${\cal P}(V+V{\cal Q}G_{0}{\cal Q}V){\cal P}$
between one- and three-body states.
}
\label{fig5}
\end{figure}

\begin{figure}[h]
\begin{center}
\epsfile{file=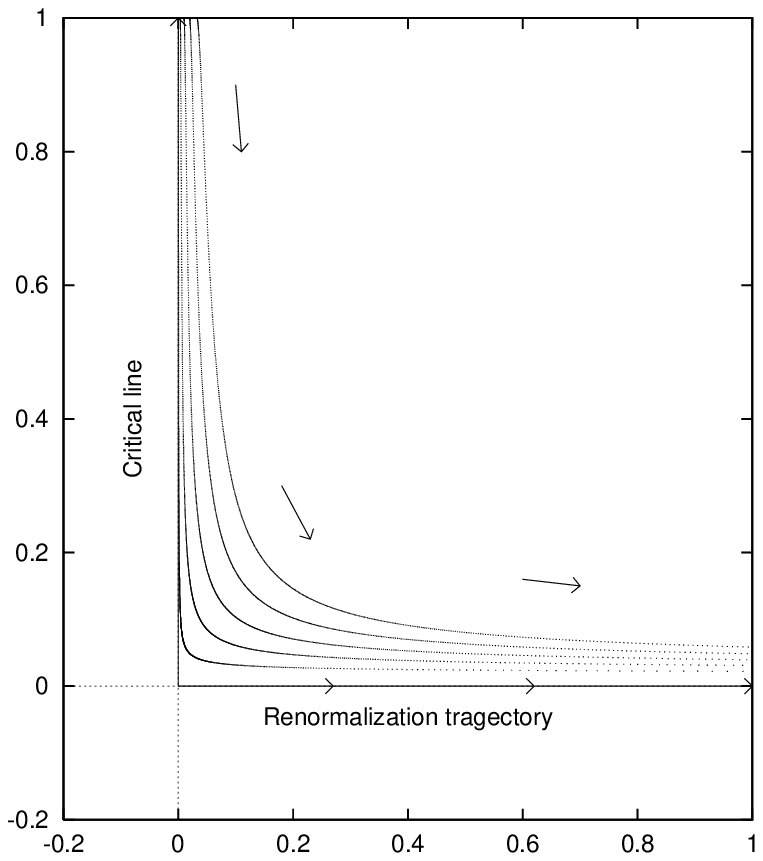,scale=1.2}
\vspace{1.0cm}
\put(-150,-15){$\mu'^2/\Lambda_0^2$}
\put(-300,150){$\lambda'$}
\end{center}
\caption{
 A flow diagram for the lowest eigenstate with $M_1/\Lambda_0=0.01$.
The vertical and horizontal axes mean
$\mu'^2$ and $\lambda'$, respectively.
The first quadrant($\lambda'>0$, $\mu'^2>0$)
corresponds to the symmetric phase.
The renormalization trajectory exists on the positive $\mu'^2$
axis and the critical line on the positive $\lambda'$ axis.
Flows are plotted with initial conditions,
$\mu'_0/\Lambda_0=0.2, 0.6, 1.0, 1.4, 1.8$ and $\lambda'_0=1.0$,
and with a step $\delta\Lambda/\Lambda_0=0.001$.
}
\label{fig6}
\end{figure}
\begin{figure}[h]
\begin{center}
\epsfile{file=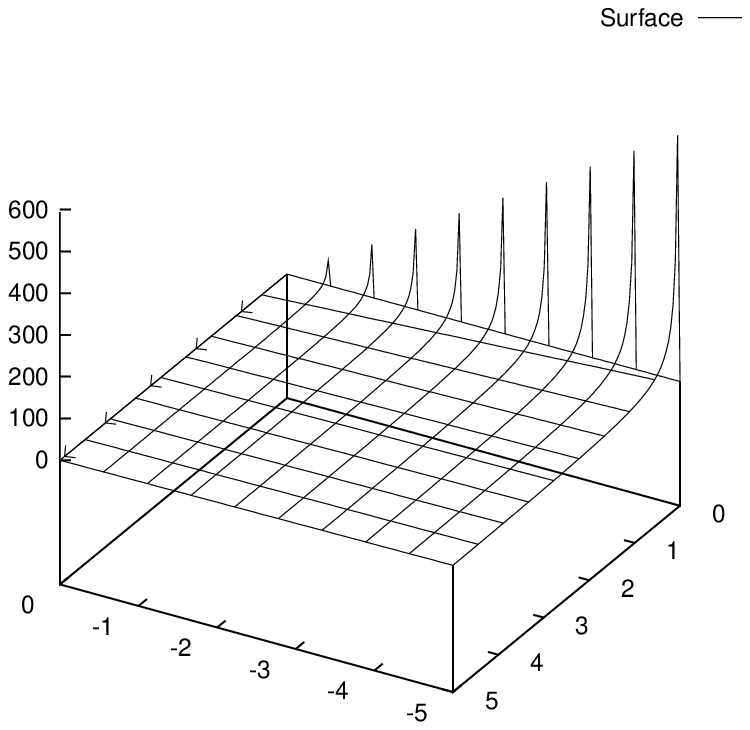,scale=1.2}
\put(-60,50){$v^2$}
\put(-210,30){$\mu^2$}
\put(-300,150){$\lambda$}
\put(-230,165){\rm R.T.}
\end{center}
\caption
{
A flow diagram for the lowest eigenstate is on the surface
$\mu^2+v^2\lambda/6=0$ corresponding to the broken phase.
The renormalization trajectory(R.T.) is on the positive $v^2$ axis.
}
\label{fig7}
\end{figure}
\end{figure}

\begin{thebibliography}{99}
  \bibitem{phw}
   R. J. Perry, A. Harindranath and K. G. Wilson,
   Phys. Rev. Lett {\bf 65}, 2959(1990).
  \bibitem{tam}
   I. Tamm,
   J. Phys. (USSR){\bf 9}, 449(1945);
   S. M. Dancoff,
   Phys. Rev. {\bf 78}, 382(1950);
   H. A. Bethe and F. D. Hoffman,
   {\it Mesons and Fields}(Row, Peterson, Evanson, 1955)Vol. II;
   E. M. Henley and W. Thirring,
   {\it Elementary Quantum Field Theory}(McGraw-Hill, New York, 1962)
  \bibitem{hara}
   K. Harada, T. Sugihara, M. Taniguchi and M. Yahiro,
   Phys. Rev. {\bf D49}, 4226(1994).
  \bibitem{sugi}
   T. Sugihara, M. Matsuzaki and M. Yahiro,
   Phys. Rev. {\bf D}, to be published.
  \bibitem{my}
   T. Maskawa and K. Yamawaki,
   Prog. Theor. Phys. {\bf 56}, 270(1976);
  \bibitem{ps}
   S. S. Pinsky and B. van de Sande,
   Phys. Rev. {\bf D49}, 2001(1994).
  \bibitem{mpsw}
   D. Mustaki, S. S. Pinsky, J. Shigemitsu and K. G. Wilson,
   Phys. Rev. {\bf D43}, 3411(1991)
  \bibitem{atw}
   O. Abe, K. Tanaka and K. G. Wilson,
   Phys. Rev. {\bf D48}, 4856(1993)
  \bibitem{ghpsw}
   St. G{\l}azek, A. Harindranath, S. S. Pinsky, J. Shigemitsu
   and K. G. Wilson,
   Phys. Rev. {\bf D47}, 1599(1993).
  \bibitem{wk}
   K. G. Wilson and J. Kogut
   Phys. Rep. {\bf C12}, 75(1974).
  \bibitem{per1}
   R. J. Perry
   Ann. Phys. (N.Y.){\bf 232}, 116(1994)
  \bibitem{wil1}
   K. G. Wilson and J. Kogut
   Phys. Rev. {\bf D8}, 1438(1970).
  \bibitem{gw}
   St. G{\l}azek and K.G. Wilson,
   Phys. Rev. {\bf D48}, 5863(1993);
   Phys. Rev. {\bf D49}, 4214(1993).
  \bibitem{b}
   C. Bloch,
   Nucl. Phys. {\bf 6}, 328(1958).
  \bibitem{bh}
   C. Bloch and J. Horowitz,
   Nucl. Phys. {\bf 8}, 91(1958).
  \bibitem{chang}
   S. -J. Chang, R. G. Root and T. -M. Yan,
   Phys. Rev. {\bf D7}, 1133(1973).
   S. -J. Chang and T. -M. Yan,
   Phys. Rev. {\bf D7}, 1147(1973).
   T. -M. Yan,
   Phys. Rev. {\bf D7}, 1760(1973);
   Phys. Rev. {\bf D7}, 1781(1973).
  \bibitem{ny}
   N. Nakanishi and K. Yamawaki,
   Nucl. Phys. {\bf B122}, 15(1977).
  \bibitem{pw}
   P. J. Perry and K.G. Wilson,
   Nucl. Phys. {\bf B403}, 587(1993)
  \bibitem{ohio}
   K. G. Wilson, T. S. Walhout,
   A. Harindranath, W. M. Zhang, R. J. Perry and
   S.t G{\l}azek, Phys. Rev. {\bf D49}, 6720(1994).
  \bibitem{bps}
   C. M. Bender, S. S. Pinsky and B. van de Sande,
   Phys. Rev. {\bf D48}, 816(1993).
  \bibitem{bsh}
   S. S. Pinsky, B. van de Sande and J. R. Hiller
   Phys. Rev. {\bf D51}, 726(1995).
\end{thebibliography}
\end{document}